\documentclass[11pt]{article}
\usepackage{soul}
\usepackage{amssymb}
\usepackage{url}
\usepackage{amsfonts}
\usepackage{amsmath}
\usepackage[misc,geometry]{ifsym}
\usepackage[top=1.2in, bottom=1.2in, left=1.2in, right=1.2in]{geometry}
\usepackage{graphicx,color,psfrag}
\usepackage[latin1]{inputenc}
\usepackage{multirow}
\usepackage{rotating}
\usepackage{natbib}
\usepackage{threeparttable}
\usepackage{booktabs}
\usepackage{longtable}
\usepackage{titlesec}
\usepackage{array}
\usepackage{tikz}
\usepackage{hyperref}
\usepackage{enumitem}
\usepackage{titling}
\usepackage{bbm}
\usepackage{titling}
\usepackage{rotating}
\usepackage{authblk}
\usepackage{soul}
\usepackage{caption}
\usepackage{subcaption}
\usepackage{float}

\usepackage{pdflscape}

\hypersetup{
	pdftitle={},    
	pdfauthor={},     
	colorlinks=true,       
	linkcolor=blue,          
	citecolor=black,        
	filecolor=black,      
	urlcolor=black           
}
 
\newcolumntype{L}[1]{>{\raggedright\let\newline\\\arraybackslash\hspace{0pt}}m{#1}}
\newcolumntype{C}[1]{>{\centering\let\newline\\\arraybackslash\hspace{0pt}}m{#1}}
\newcolumntype{R}[1]{>{\raggedleft\let\newline\\\arraybackslash\hspace{0pt}}m{#1}}

\titleformat*{\section}{\centering\Large\sc}
\titleformat*{\subsection}{\large\bf}

\renewcommand{\baselinestretch}{1.3} 

\begin{document}
\title{The 15-Minute City Quantified Using Mobility Data \thanks{These authors contributed equally. Corresponding Author: \Letter \,\,ariana@mit.edu.}}
\author[1]{Timur Abbiasov*}
\author[1]{Cate Heine*}
\author[2]{Edward Glaeser*}
\author[1]{Carlo Ratti}
\author[1]{Sadegh Sabouri*}
\author[1\,{\Letter}]{Arianna Salazar Miranda*}
\author[1]{Paolo Santi}

\affil[1]{Senseable City Lab, MIT}
\affil[2]{Harvard University}
\date{November 2022}
\maketitle

\begin{abstract}
{\footnotesize 
Americans travel 7 to 9 miles on average for shopping and recreational activities, which is far longer than the 15-minute (walking) city advocated by ecologically-oriented urban planners. This paper provides a comprehensive analysis of local trip behavior in US cities using GPS data on individual trips from 40 million mobile devices. We define \emph{local usage} as the share of trips made within 15-minutes walking distance from home, and find that the median US city resident makes only 12\% of their daily trips within such a short distance. We find that differences in access to local services can explain eighty percent of the variation in 15-minute usage across metropolitan areas and 74 percent of the variation in usage within metropolitan areas. Differences in historic zoning permissiveness within New York suggest a causal link between access and usage, and that less restrictive zoning rules, such as permitting more mixed-use development, would lead to shorter travel times. Finally, we document a strong correlation between \emph{local usage} and experienced segregation for poorer, but not richer, urbanites, which suggests that 15-minute cities may also exacerbate the social isolation of marginalized communities.
}
\end{abstract}

\thispagestyle{empty}
\setcounter{page}{0}
\clearpage

\section{Introduction}
Do Americans travel relatively long distances for shopping, personal business, and recreational activities because they lack access to nearby amenities? The National Household Travel Survey (NHTS) reports that in 2017, the average shopping trip was 7.3 miles, the average trip for personal business, such as pet care, was 6.8 miles and the average trip for recreational activities, such as visiting the gym, was 8.7 miles. These distances are far from the 15-minute (walking) city advocated by urban planners who hope that shorter trips will reduce carbon emissions and traffic jams. If those long trips occur because of an absence of nearby options, as opposed to idiosyncratic tastes or a love of variety, then urban planning and mixed-use zoning could potentially reduce trip lengths significantly, but shorter trips could also mean a more segregated metropolitan area. In this paper, we estimate the relationship between trip length and access to nearby amenities and explore whether shorter trip length leads to more experienced segregation.

We use cell phone data to measure trip lengths across the US. While cell phone records lack the detailed household information that is available on the National Household Travel Survey, the sample contains 11 billion point of interest (POI) visits in 2019. The richness of the data enables us to measure mobility patterns throughout the US, and to relate those patterns to features of the local economic geography, especially the proximity of the points of interest to residences \citep{Ratti2006, Jiang2012, becker2013human, Gonzalez2008, schlapfer_2021}. We also measure experienced segregation with these cell-phone records (following \cite{wang2018urban, xu2019,athey_2021,moro_2021}) and test whether experienced segregation increases with nearby access and shorter trips.

The cell phone data identifies each subject's home census block group, and we focus on the share of trips to restaurants, schools, parks, healthcare, drugstores, arts and cultural institutions, grocery stores, services, and religious organizations, that are within a 15-minute walk of that block group. To compute this measure of local usage, we scrape 15-minute walksheds around the centroid of every home census block group following the street network. We focus on 15 minutes because of the recent policy interest in 15-minute cities, but we recognize that any particular number is relatively arbitrary.\footnote{Appendix Table \ref{tbl:corr cut-offs} reports the high correlation between the 15-minute measure and others based on the share of trips within 10, 20, or 25 minutes, and measures based on median trip distance.} We focus on those particular points of interest because they represent the focus on much of the policy discussion surrounding 15-minute cities.

After discussing our methods in Section \ref{measurement}, Section \ref{sec: us overview} describes the basic patterns of trip length across the US. Only twelve percent of trips in the average urban census block group lead to locations that are within a 15-minute walk of the block group, but there is significant regional variation. In the northeast, thirty-two percent of urban trips involve points of interests that are within 15 minutes of the home census block group. That 15-minute share falls to 16 percent in the south and it is still lower in most of the west. Income and education are both associated with longer trips, which is also well documented in the NHTS. Population density and the share of households in the block group without a car are both negatively associated with the share of trips beyond the 15-minute walkshed.

To test the connection between use of nearby services and access to nearby services, in Section \ref{sec: correlates}, we form an index of access based on the point of interest data (following, for example, \cite{weng2019,capasso2019, gaxiola2021, graells2021, calafiore2022}). For each of the eight different categories, we count the number of POIs within a 15-minute walk of the census block group. We then take block groups across urban areas within each category, and assign the block group a category-specific number based on the share of other block groups that have less nearby access to that category of amenity. We then average across those eight categories to get an index between 0 and 100 roughly capturing the share of block groups that have less 15-minute access.

In a statistical sense, 15-minute access can explain eighty percent of the variation in 15-minute usage across metropolitan areas and 74 percent of the variation in usage within metropolitan areas. A one percentile increase in access is associated with a .8 percentage point increase in the share of trips that are within 15 minutes walking distance. This coefficient is not particularly sensitive to other controls, including population density, income and share of the population in the block group that owns a car. This correlation does not imply that encouraging more mixed-use development within residential areas will reduce average trip times, but it is certainly compatible with that hypothesis.

There are two central confounds to a causal interpretation of the cross-section correlation between access to nearby amenities and short trips to those amenities. First, businesses may locate in places where people want to walk to nearby shops. Second, people who like walking to nearby service providers may choose to locate in places that have abundant service-providers nearby.

In section \ref{sec: causal}, we address the first concern by following \citet{martynov2022} and \citet{shertzer2018} in using the variation created by 1961 New York City Zoning ordinance across New York Neighborhoods. We focus on the average maximum commercial floor area ratio (FAR) allowed in each neighborhood by the plan. This variable strongly predicts 15-minute access in 2019. When we use this variable as an instrument for access within New York City, the estimated coefficient when 15-minute usage is regressed on 15-minute access becomes somewhat stronger. We interpret this result as both suggesting that the endogenous elements of firm location choice is not driving our result, and making the case that local zoning rules certainly do shape the level of access to nearby amenities.

In Section \ref{sec: social disparities}, we test the hypothesis that fomenting 15 minute access and usage could increase the level of segregation in the city \citep{glaeser_2021, velts_2022}. We first regress experienced segregation on 15-minute usage and the interaction between 15-minute usage and income. We find that 15-minute usage is strongly positively correlated with experienced segregation for the poor and negatively correlated with experienced segregation for the rich. We do the same thing looking at 15-minute access and find analogous results. These results support the view that, while improving access to amenities reduces trip lengths, it also poses risks of exacerbating the social isolation of marginalized communities.

\section{Measuring 15-minute usage and access}\label{measurement}

\paragraph{Measuring 15-minute usage:} We use large-scale mobile data tracing everyday trips for approximately 40 million smartphone users in the 418 most populated urban areas in the US. The data is provided by SafeGraph and contains information on trips to 5.3 million points of interest (POIs) across the country (3.9 million of which are in our sample urban areas). This dataset provides the location of each POI (latitude and longitude), categorizes it according to the North American Industry Classification System (NAICS) code, and describes how many individuals visited the POI from each home census block group in the US. In total, the dataset covers 13.6 billion POI visits over the course of 2019---our study period.

To create our measure of 15-minute usage, we focus on amenities that represent the types of essential functions associated with the 15-minute city model \citep{moreno_2021}: restaurants, schools, parks, healthcare, drugstores, arts and culture institutions, grocery stores, services, and religious organizations.  These together make up 30\% of the POIs in our dataset and account for 35\% of the trips in our data. Our 15-minute usage measure is defined as the share of trips originating in each neighborhood that occur within a 15-minute walking distance from that block group's centroid.

\paragraph{Measuring access:} To assess the relationship between use of nearby services and access to nearby services, we compute a measure of access that describes the number of essential amenities reachable within a 15-minute walking distance from one's home census block group.

As with our measure of usage, we focus on amenities in the eight categories introduced above. For each category, we count the number of POIs within a 15-minute walk of the census block group and assign each neighborhood a score given by the fraction of neighborhoods with fewer POIs in that category. We average these scores for all categories (weighting by the average share of trip in each at the national level) to get an index between 0 and 100 of local access.\footnote{Table \ref{tbl:corr cut-offs} shows the correlations between our baseline access metric and alternative measures of access including share of trips within 10, 20 and 25 minutes, and median trip distance. These alternative measures of access are highly correlated across neighborhoods. While 15-minutes may be a relatively arbitrary number, our results are quite similar if we use other numbers or similar measures.}

\begin{figure}[t]    
    \centering
        \centering  
        \includegraphics[width=\textwidth]{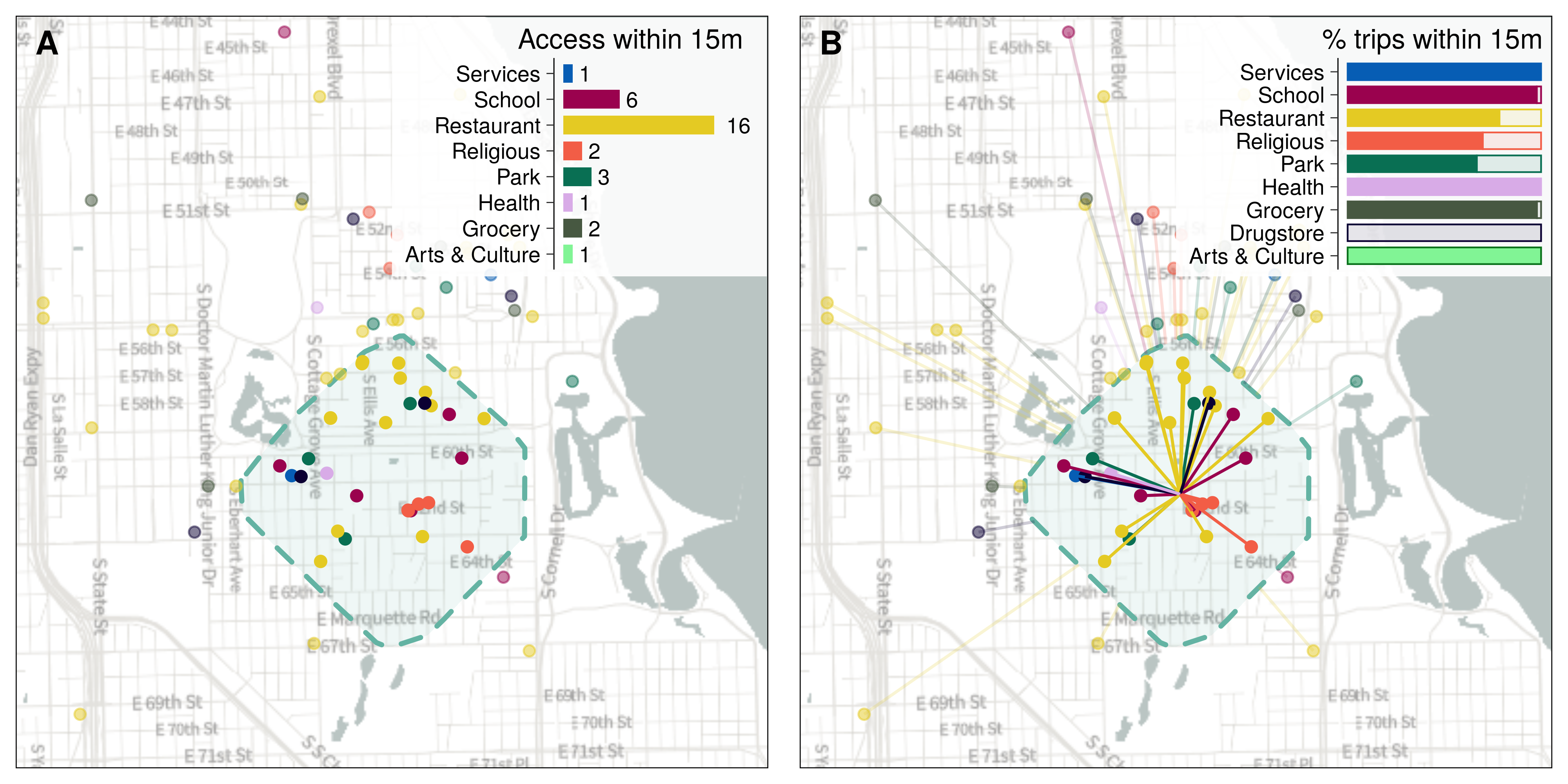}
        \caption{\textsc{Example of Measuring access and usage within a 15-minute walk  with Safegraph data.} Panel A displays the POIs within a 15-minute walkshed from the centroid of the neighborhood. Panel B displays trips to amenities within and outside the 15-minute walkshed and reports the share of trips within the 15-minute walkshed by type of POI in the upper-right corner.}
        \label{fig: isomap}     
\end{figure}

Figure \ref{fig: isomap} provides an example of how we constructed our measure of access (left panel) and 15-minute usage (right panel) for a neighborhood in Chicago. The access measure captures the availability of amenities within a 15 minute walking radius (depicted by the dashed green boundary) from the centroid of the neighborhood. The 15-minute usage measure is given by the number of trips from that neighborhood to POIs within the 15-minute walkshed as a share of all trips originating in that neighborhood.

Our main analysis is conducted at the neighborhood level (identified as census block groups). We also provide analyses aggregating the data at the urban area level, weighting each neighborhood by its population.    

\section{Results}\label{sec: us overview}
\subsection{Documenting local trip behavior in the US}

\begin{figure}[t]
	\centering
        \includegraphics[width=1\linewidth]{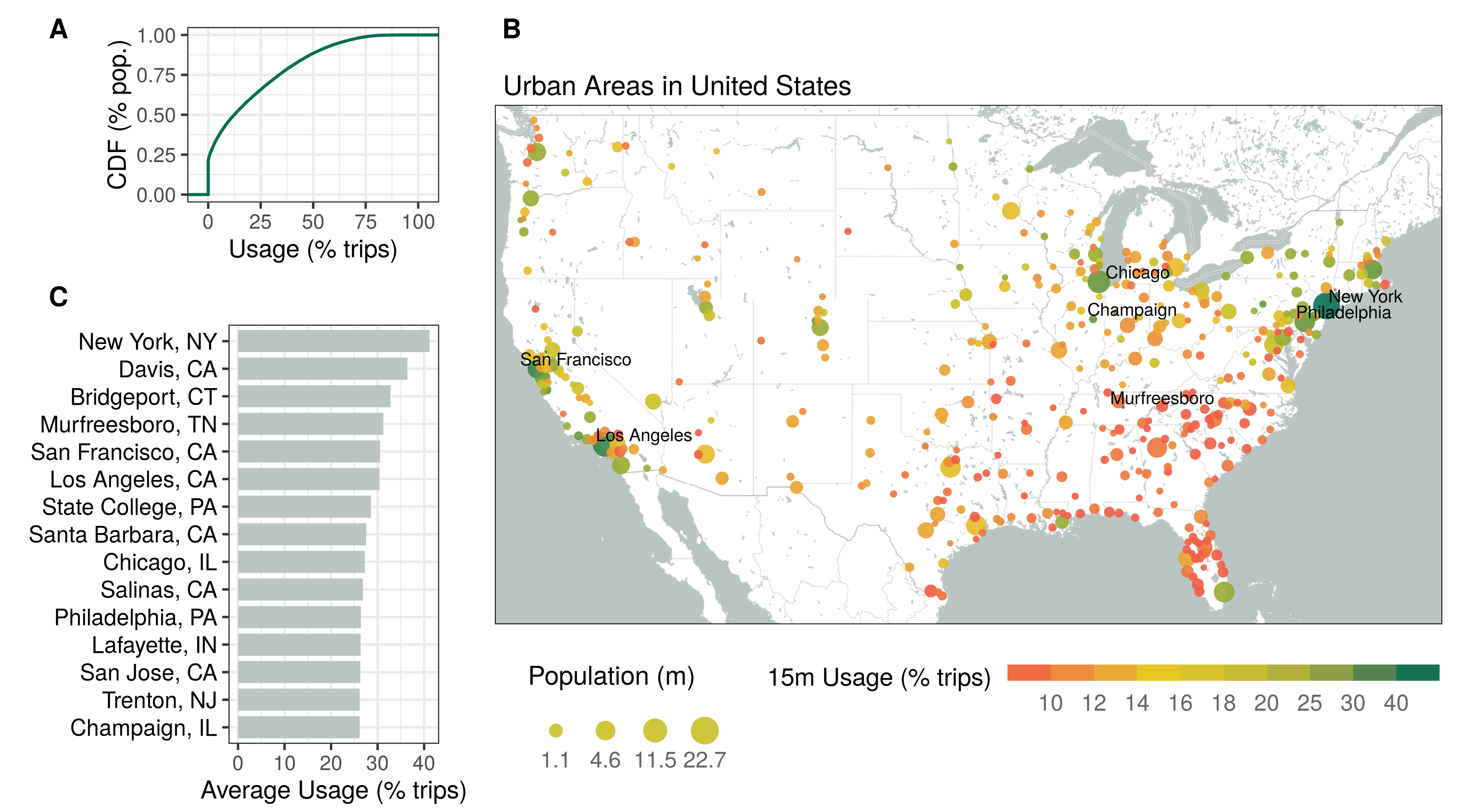}
	\caption{\textsc{Local trips in the US.} Panel A plots the cumulative distribution of 15-minute usage for US neighborhoods. Panel B plots 15-minute usage for all urban areas in the US. Panel C reports 15-minute usage for urban areas with the highest shares of local trips. }
	\label{fig: usage map}
\end{figure}

We first document the level of 15-minute usage across the US. Panel A in Figure \ref{fig: usage map} plots the cumulative distribution of our measure of 15-minute usage across all US neighborhoods. Residents of the median US neighborhood only take 12.1\% of their trips to basic amenities within a 15-minute walking radius. Residents in 87\% of US neighborhoods make at least half their trips outside a 15-minute walking radius.

Panel B in Figure \ref{fig: usage map} plots 15-minute usage for the median resident in each US urban area. The map depicts a clear regional divide, with residents of urban areas in the Northeast taking close to 32\% of all trips within a 15-minute walking distance. In comparison, residents of Southern urban areas take 16\% or fewer of all trips within 15-minute walking distance, where people rely more heavily on personal vehicles to conduct their local trips \citep{FHWA2017}.

Panel C in Figure \ref{fig: usage map} shows the average 15-minute usage measure for urban areas with the highest share of local trips. For example, 47.7\% of daily trips fall within a 15-minute walk from home for the residents of New York.

\begin{figure}[t]
    \centering
        \centering
        \includegraphics[width=0.7\textwidth]{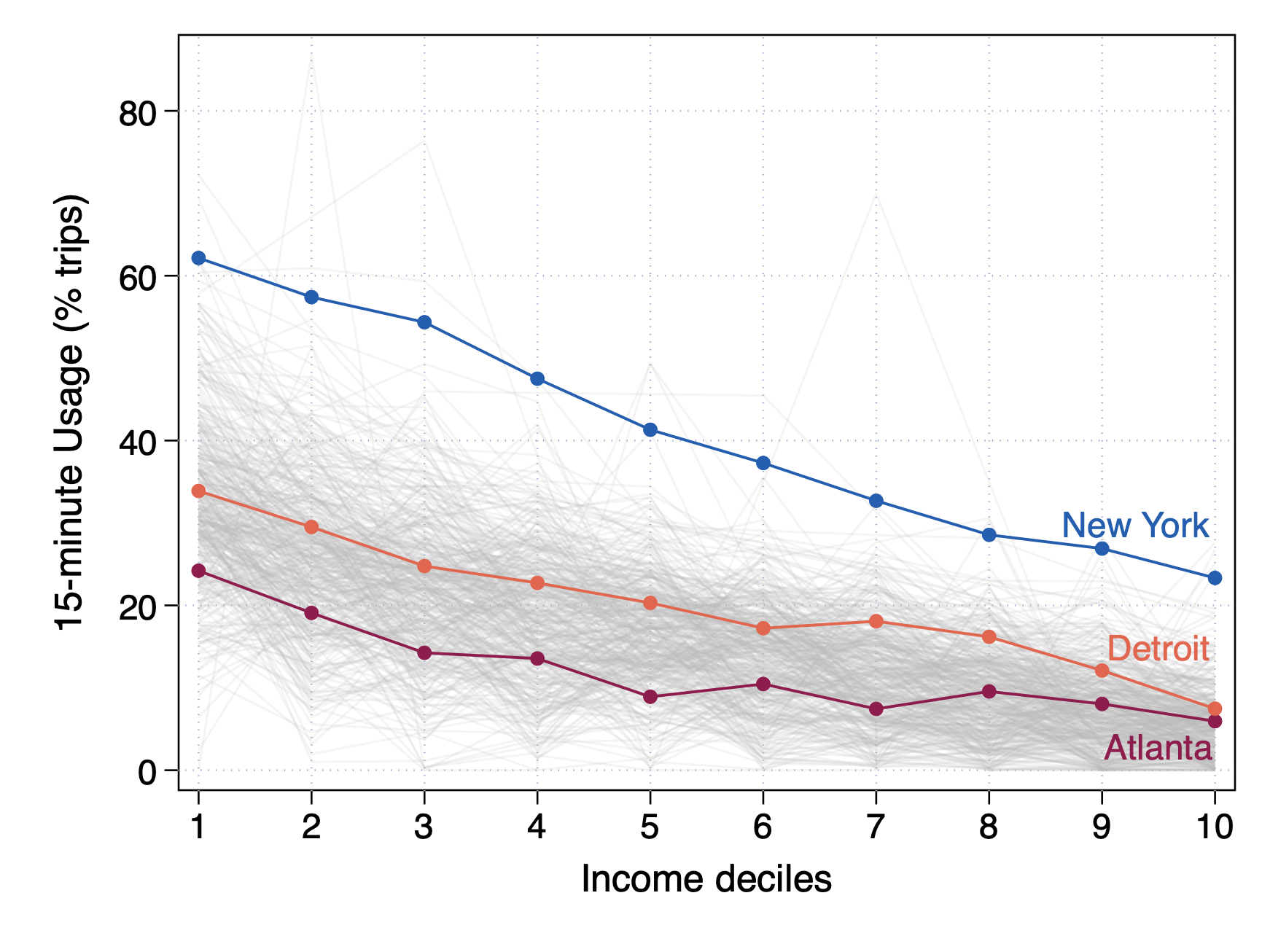}
        \caption{\textsc{Local trips by income levels.} The figure plots 15-minute usage by neighborhood income deciles for all urban areas, including New York, Detroit, and Atlanta. 
 }
	\label{fig: usage access by income}
\end{figure}

While we do not have attributes of the individuals themselves, Census data provides us with attributes of their neighborhoods. Figure \ref{fig: usage access by income} plots 15-minute usage by neighborhood income deciles for all urban areas. We highlight New York, Detroit, and Atlanta, as examples of places with particularly high, medium, and low levels of local usage. In every urban area, 15-minute usage decreases sharply with income. Residents of the neighborhoods in the bottom income decile make 51\% of their trips within a 15-minute radius, while residents of the richest 10\% only make 17\% of their trips locally. 

\subsection{The role of access in local trips}\label{sec: correlates}

\begin{figure}[t]
    \centering
        \includegraphics[width=1\linewidth]{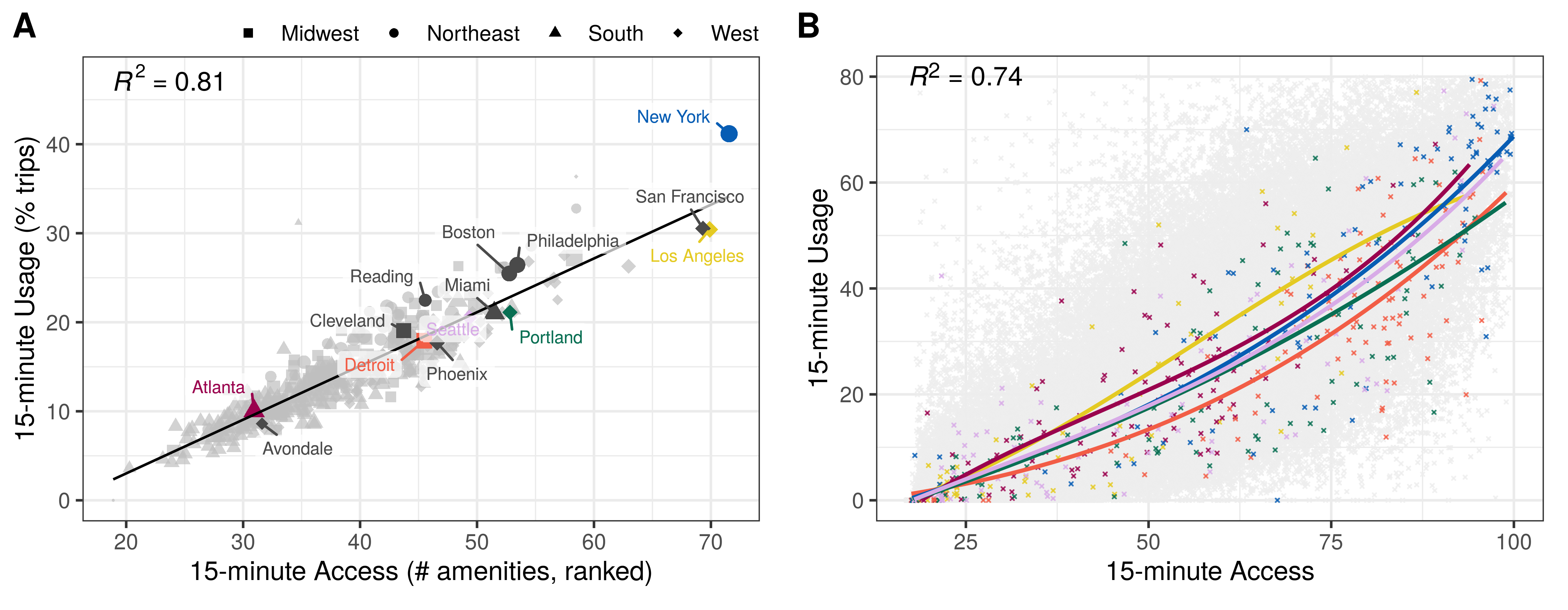}
        \caption{\textsc{Access and local trips.} Panel A plots the relationship between 15-minute usage and access at the urban area level. Panel B plots the relationship between 15-minute usage and access across neighborhoods within urban areas, and highlights block groups in six example urban areas, with colors corresponding to cities highlighted in Panel A. For each of the six urban areas, Panel B also plots a fitted non-parametric spline regression using that same color. In Panel A, the R-squared is obtained from a regression of 15-minute usage on 15-minute access at the urban area level. In Panel B, the R-squared is from a regression at the census block group level with urban area fixed effects.}
        \label{fig: usage access}
\end{figure}

Planners who advocate 15-minute cities focus on providing access to amenities, which they hope will lead to more local trips. We now turn to the relationship between access to local amenities and 15-minute usage. Figure \ref{fig: usage access}, Panel A plots 15-minute usage against our measure of access aggregated at the urban area level. 15-minute access explains 81\% of the variation in 15-minute usage across urban areas. Figure \ref{fig: usage access}, Panel B plots the relationship between 15-minute usage and access across neighborhoods. Even within urban areas, 15-minute usage rises sharply with access. The level of 15-minute usage increases from 0 to roughly 60\% as we move from the neighborhoods with the lowest local access to neighborhoods with the highest local access. Differences in access account for 74\% of the within-city variation in local usage across neighborhoods.   

\begin{table}[!ht] \centering 
  \caption{Relationship between 15-minute usage and access across neighborhoods (Census Block Groups) and Urban areas} 
 \label{tbl: main covariates}
\resizebox{\textwidth}{!}{\begin{tabular}{@{\extracolsep{5pt}}lcccccccc} 
\toprule
 & \multicolumn{8}{c}{\textsl{Dependent Variable:} 15-minute Usage (\% of trips)} \\
 & \multicolumn{4}{c}{Across Neighborhoods} & \multicolumn{4}{c}{Across Urban Areas} \\ \cmidrule(r){2-5} \cmidrule(r){6-9}
\\[-1.8ex] & (1) & (2) & (3) & (4) & (5) & (6) & (7) & (8)\\ \midrule
 15-minute Access & 0.761$^{***}$ & 0.696$^{***}$ & 0.686$^{***}$ & 0.693$^{***}$ & 0.604$^{***}$ & 0.481$^{***}$ & 0.426$^{***}$ & 0.424$^{***}$ \\ 
  & (0.006) & (0.007) & (0.007) & (0.007) & (0.015) & (0.035) & (0.031) & (0.031) \\ 
  & & & & & & & & \\ 
 Pop Density Log &  & 0.787$^{***}$ & 0.225 & 0.473$^{***}$ &  & 2.473$^{***}$ & 2.308$^{***}$ & 2.471$^{***}$ \\ 
  &  & (0.265) & (0.174) & (0.165) &  & (0.634) & (0.564) & (0.576) \\ 
  & & & & & & & & \\ 
 Median Income Log &  & $-$4.282$^{***}$ & $-$1.954$^{***}$ & $-$1.583$^{***}$ &  & $-$2.989$^{***}$ & $-$0.171 & $-$0.340 \\ 
  &  & (0.239) & (0.210) & (0.231) &  & (0.703) & (0.680) & (0.704) \\ 
  & & & & & & & & \\ 
 \% Education: BA + &  & 0.008 & $-$0.016$^{***}$ & $-$0.012$^{***}$ &  & 0.050$^{***}$ & 0.034$^{***}$ & 0.029$^{**}$ \\ 
  &  & (0.005) & (0.004) & (0.004) &  & (0.014) & (0.012) & (0.013) \\ 
  & & & & & & & & \\ 
 \% White &  & $-$0.007 & 0.008 & 0.006 &  & $-$0.002 & 0.017$^{*}$ & 0.014 \\ 
  &  & (0.005) & (0.005) & (0.005) &  & (0.010) & (0.009) & (0.009) \\ 
  & & & & & & & & \\ 
 \% No-car Households &  &  & 0.175$^{***}$ & 0.176$^{***}$ &  &  & 0.483$^{***}$ & 0.464$^{***}$ \\ 
  &  &  & (0.009) & (0.009) &  &  & (0.046) & (0.051) \\ 
  & & & & & & & & \\ 
 Transit Density Log &  &  &  & 6.323$^{**}$ &  &  &  & $-$1.651 \\ 
  &  &  &  & (2.934) &  &  &  & (6.352) \\ 
  & & & & & & & & \\ 
 Average Commute (mi) &  &  &  & 0.209$^{***}$ &  &  &  & $-$0.018 \\ 
  &  &  &  & (0.013) &  &  &  & (0.011) \\ 
  & & & & & & & & \\ 
Urban Area FE & \checkmark & \checkmark & \checkmark & \checkmark &  &  &  &  \\ 
Observations & 147,760 & 143,417 & 143,417 & 143,401 & 418 & 418 & 418 & 418 \\ 
R$^{2}$ & 0.736 & 0.752 & 0.759 & 0.761 & 0.806 & 0.819 & 0.857 & 0.858 \\ 
\bottomrule
\end{tabular}} 
\begin{minipage}{1\linewidth}
\scriptsize \textsl{Notes.---} The table reports OLS estimates of the relationship between 15-minute usage and access.
Columns 1 to 4 report estimates at the neighborhood level (controlling for urban area fixed effects) and columns 5 to 8 report estimates at the urban area level. The covariates included in columns 2 to 4 and 6 to 8 come from the American Community Survey and the \emph{National Transit Map} from the Bureau of Transportation Statistics.    
Robust standard errors clustered at the county level are in parentheses. The coefficients with *** are significant at the 1\% confidence level; with ** are significant at the 5\% confidence level; and with * are significant at the 10\% confidence level.					
\end{minipage}

\end{table}

Table \ref{tbl: main covariates} reports results from regressions of 15-minute usage on the measure of access estimated across neighborhoods (columns 1-4) and across urban areas (columns 5-8). Column 1 reports that a one percentile increase in access is associated with a .76 percentage point increase in 15-minutes usage.  The coefficient falls to approximately .69 when we control for neighborhood demographics (column 2), car ownership (column 3), and the density of public transit, and commutes lengths (column 4).  Neighborhood income and education are associated with longer trips, which corresponds to the well-known correlation between trip length and household income in the NHTS \citep{pucher_2003}. Population density and the share of households in the block group without a car are both negatively associated with the share of trips beyond the 15-minute walkshed. Columns 5-8 confirm that there is also a strong link between 15-minute access and 15-minute usage across urban areas. Differences in access can explain 80\% of the variation in 15-minute usage across cities.

\subsection{Towards the causal effect of access on 15-minute usage for New York City}\label{sec: causal}

The strong positive correlation described above supports the hypothesis that encouraging more mixed-use development within residential areas will reduce average trip times. However, the strength of this correlation may not reflect the causal effect of improving access on local trips. Higher use of local amenities may attract new businesses and developers to the area, so that the regression reflects the influence of usage on access rather than the reverse. Robust access to local amenities may attract residents who like to take shorter trips, even if access does little to induce an average person to take shorter trips.  

This section addresses the first concern by using the spatial variation in the zoning code introduced in New York City in 1961 \citep{martynov2022, twinam2017}. The 1961 zoning resolution established for the first time city-wide floor-to-area ratio (FAR) regulations, dividing New York City into 1196 zoning districts, where different maximal densities for residential, commercial, and other types of uses were permitted. Figure \ref{fig: zoning map} in the Appendix provides a map of these areas. We focus on the average maximum commercial floor area ratio (FAR) allowed in each neighborhood and use it to provide exogenous variation in current access to amenities and services.

This source of variation should solve the first but not the second problem discussed above, as long as the variation in business density induced by the 1961 code does not reflect any tendency of the tastes of current residents to attract businesses.   The 1961 zoning plan in New York city was primarily driven by past built-up densities and broad geographic definitions. Consequently, we follow \citet{shertzer2018} and include controls for population and built-up density, proximity to the city center, employment composition, poverty, and the share of non-white population, obtained from the 1960 census data.

Relying on the 1961 code for identification will not, however, fix the problem associated with the selection of people who like short trips into neighborhoods with abundant local access.  Even if the characteristics of retail in a neighborhood are completely random, residents will still sort into neighborhoods that fit their needs.  Consequently, we view our instrumental variables estimates as a step towards causal inference rather than perfect estimates of the causal effect of access on usage.   

\begin{figure}[t]
	\centering
        \includegraphics[width=1\linewidth]{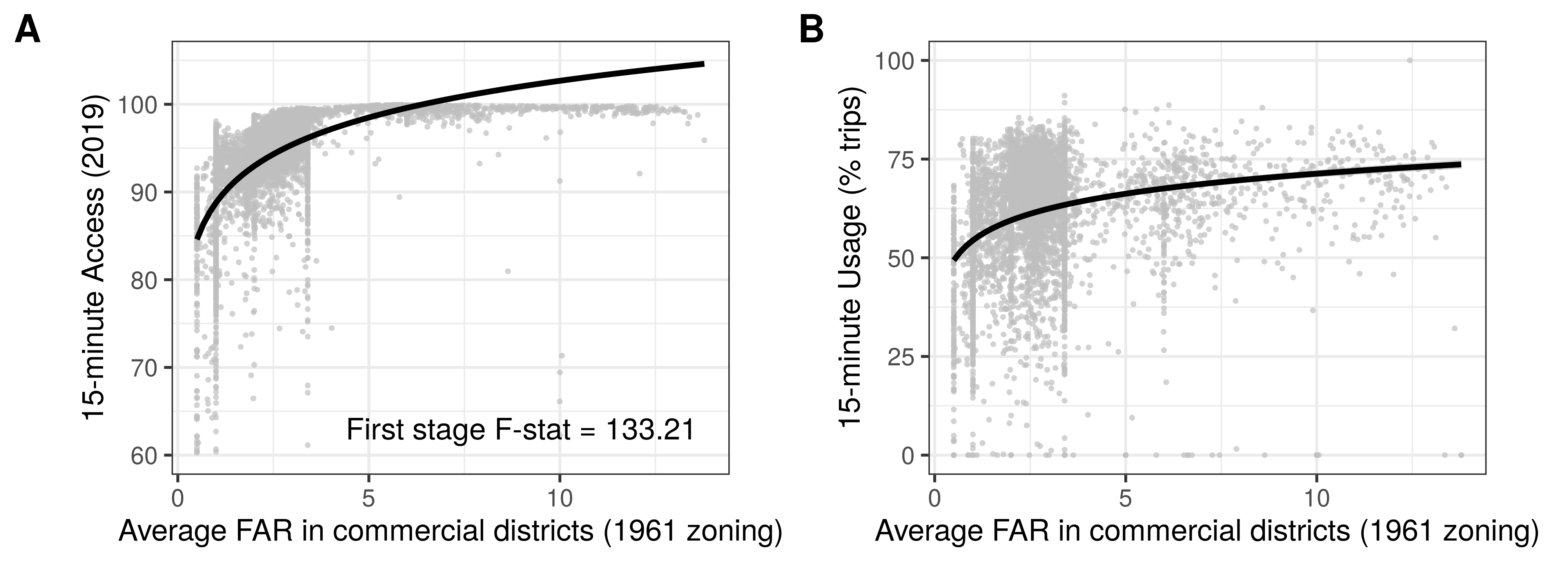}
	\caption{\textsc{First-stage and reduced-form relationships between historical zoning regulations and current access to local services and 15-minute usage for New York neighborhoods.} Panel A shows the first-stage relation between historical zoning and current access. Panel B shows the reduced-form relationship between historical zoning and current 15-minute usage. Table \ref{regs_far_usage_access} in the Appendix summarizes these results.}
	\label{fig: ivregs summary}
\end{figure}

Panel A in Figure \ref{fig: ivregs summary} reports the first-stage relationship between average FAR allowed by the 1961 zoning resolution and our measure of access across neighborhoods in New York. There is a strong first-stage association, with an F-statistic of 370, indicating that neighborhoods with fewer FAR restrictions in surrounding commercial districts see higher contemporary levels of access to amenities within a 15-minute walking distance. Panel B reports the reduced-form relationship between historical FAR allowances and 15-minute usage. We also see here that residents of neighborhoods with less historical restrictions on commercial areas take short trips more frequently, as captured by an increase in our 15-minute usage measure from 50\% to 75\%.

Table \ref{tbl: iv access usage} reports our IV estimates for New York City. Column 1 reports OLS estimates of the relationship between 15-minute usage and access for New York City. This estimated .88 coefficient is lower than the corresponding coefficients estimated in Table \ref{tbl: iv access usage}, columns (2)-(4).  The range of access in New York City is smaller than in our full range of urban areas, which may explain this smaller coefficient. Columns (2)-(4) provide our IV estimates,  controlling for different covariates.  Columns (3) and (4) provide our preferred specification and they suggest that a one percentile increase in access is associated with a 1.1 percentage point increase in local usage.  This estimate is closer to the national OLS estimates, although larger than its New York City OLS counterpart, which may reflect attenuation in the OLS estimates due to measurement error.\footnote{Given that we only have a single snapshot of data to capture available amenities in 2019, our access measurement may not perfectly reflect amenities opened and closed throughout the span of the year, and, hence, the main independent variable is measured with error. If that is the case, our linear regression estimates would be biased towards zero.}

\begin{table}[!htbp] \centering 
  \caption{OLS and IV estimates of the relationship between 15-minute usage and access across neighborhoods (Census Block Groups) in New York City} 
  \label{tbl: iv access usage}
\resizebox{\textwidth}{!}{\begin{tabular}{@{\extracolsep{5pt}}lcccc} 
\toprule
 & \multicolumn{4}{c}{\textit{Dependent variable:}} \\ 
 & \multicolumn{4}{c}{Usage (\% of trips within a 15m walk)} \\ 
 & OLS & \multicolumn{3}{c}{IV estimates} \\ \cmidrule(r){2-2} \cmidrule(r){3-5} 
 & (1) & (2) & (3) & (4)\\\midrule
 Access to amenities within a 15m walk & 0.878$^{***}$ & 1.384$^{***}$ & 1.064$^{***}$ & 1.096$^{***}$ \\ 
  & (0.033) & (0.135) & (0.120) & (0.123) \\ 
  & & & & \\ 
 Population Density (Log) & 3.554$^{***}$ &  & 3.319$^{***}$ & 3.283$^{***}$ \\ 
  & (0.254) &  & (0.355) & (0.356) \\ 
  & & & & \\ 
 Median Income (Log) & $-$2.386$^{***}$ &  & $-$2.505$^{***}$ & $-$2.583$^{***}$ \\ 
  & (0.362) &  & (0.467) & (0.479) \\ 
  & & & & \\ 
 \% Pop. with no car & 0.146$^{***}$ &  & 0.133$^{***}$ & 0.138$^{***}$ \\ 
  & (0.011) &  & (0.014) & (0.014) \\ 
  & & & & \\ 
IV: Log Average Commercial FAR in 1961 Resolution &  & \checkmark & \checkmark & \checkmark \\ 
Distance to City Center & \checkmark & \checkmark & \checkmark & \checkmark \\ 
1960 Population and Housing Density & \checkmark & \checkmark & \checkmark & \checkmark \\ 
1960 Workers by Industry & \checkmark & \checkmark & \checkmark & \checkmark \\ 
1960 Poor and Non-white Families &  &  & & \checkmark \\ 
First stage F-stat &  & 133.21 & 150.21 & 135.58 \\ 
Observations & 4,804 & 5,170 & 4,804 & 4,804 \\ 
\bottomrule
\end{tabular}} 
	\begin{minipage}{1\linewidth}													\scriptsize \textsl{Notes.---} The table reports OLS and IV estimates of the relationship between 15-minute usage and access for neighborhoods in New York City. Column 1 reports OLS estimates of the relationship between 15-minute usage and access for New York City, controlling for distance to the city center and a host of socio-economic controls for 1960, including population and housing density and the share of workers by industry. Columns 2-4 report the IV estimates. Columns 2 and 3 also control for distance to the city center, population and housing density and the share of workers by industry. Column 4 includes additional controls for the share of poor and non-white families in 1960. The covariates included in these models come from \citet{twinam2017} and \citet{martynov2022}.    
Robust standard errors are in parentheses. The coefficients with *** are significant at the 1\% confidence level; with ** are significant at the 5\% confidence level; and with * are significant at the 10\% confidence level.
	\end{minipage}	
\end{table}

These results suggest that land-use restrictions, particularly those that limit commercial development in residential areas, may lead to longer travel distances and greater carbon use. These results mirror the findings of \citet{glaeser_2010} that that residential land use restrictions seem to increase America's carbon use by limiting the growth of America's greenest metropolitan areas, such as San Francisco and San Diego, and pushing new household formation into areas, such as Atlanta and Houston, that have high carbon footprints. The global carbon consequences of local land-use rules could presumably be incorporated into any future cost-benefit analysis of such rules.   

\subsection{Trade-off between local living and segregation}\label{sec: social disparities}

To an environmental advocate, limiting cross-city travel may seem like an unalloyed good, but reducing travel across neighborhoods could also exacerbate the social isolation of marginalized communities \citep{glaeser_2021, velts_2022}. America has a long history of racial segregation within its cities, in which land use controls and then restrictive covenants were used to prevent the movement of African Americans into white locales. A policy that leads to more local trips could reduce opportunities for exchange and interaction across racial and income groups, which might further reduce opportunities for economic mobility \citep{kain_1968, chetty2022social}. 

For each neighborhood within each city, we construct a measure of experienced segregation that captures the degree to which its residents mix with members of different income levels in their daily travels. Our construction follows \citep{xu2019} and is detailed in Appendix section \ref{app: seg measure}. 

Figure \ref{fig: psi access income} Panels A and B plot our index of experienced segregation against the 15-minute access and usage measure, respectively, for neighborhoods of different income quartiles. The figure shows that there is no strong association between 15-minute usage and experienced segregation for residents of high- and upper-middle-income neighborhoods. In fact, as local trips increase, residents of high-income neighborhoods experience more social mixing. The opposite is true for residents of low-income neighborhoods. Experienced segregation rises from 67.6 to 71.4 as we move from poor neighborhoods with the lowest local usage to the highest. This increase in experienced segregation corresponds to 0.48 standard deviations of experienced segregation across neighborhoods. The increase in experienced segregation in neighborhoods within urban areas with the highest levels of local usage is much starker. In these cities, experienced segregation of low-income residents is 1.3 standard deviations higher in the highest-usage decile than the lowest-usage decile (see Table \ref{tbl : psi usage income} in the Appendix).

\begin{figure}[t]
    \centering
        \centering
        \includegraphics[width=1\textwidth]{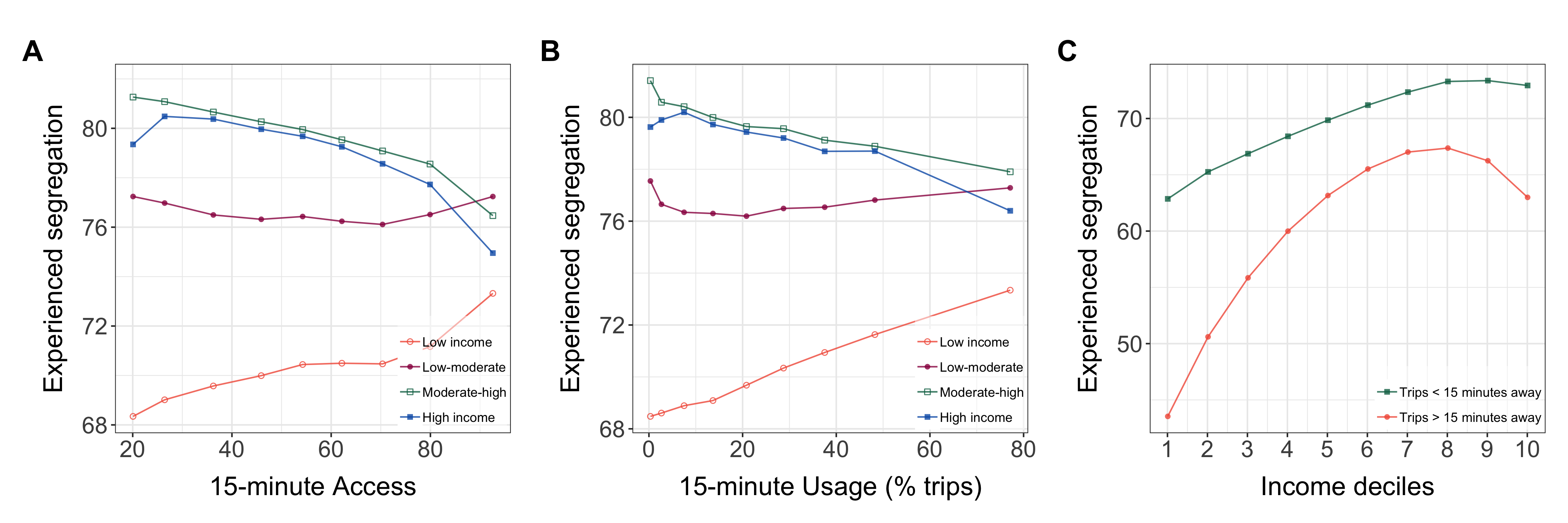}
        \caption{\textsc{Local trip behavior and experienced segregation.} Panel A plots experienced segregation against 15-minute access for neighborhoods in different income quartiles. Panel B plots experienced segregation against 15-minute usage for neighborhoods in different income quartiles. Panel C plots a measure of experienced segregation during short and long trips for residents of neighborhoods in different income deciles.
        }
        \label{fig: psi access income}
\end{figure}

The positive association between the share of 15-minute trips and our segregation index for residents of poor neighborhoods reflects the fact that people from poorer neighborhoods are more likely to encounter socioeconomically diverse individuals during long trips than shorter ones. Panel C in Figure \ref{fig: psi access income} documents this fact by plotting measures of segregation experienced during long and short trips separately for residents of neighborhoods in different deciles of the income distribution. Residents from poor households face a 29.9\% decline in experienced segregation when they travel long distances rather than short distances. Higher-income people also encounter a more diverse group of people when they leave their 15-minute walkshed than when they stay close to home, but there is only a 13.8\% decrease in experienced segregation for this group. This difference explains why the trade-off between local living and segregation is real for the poor, but not for the rich.

Table \ref{a6} in the appendix conducts a regression analysis and shows that the differences in experienced segregation for residents of poor neighborhoods documented in Figure \ref{fig: psi access income} are statistically significant. These differences are also robust to controlling for urban area fixed effects and are similar if we regress our measure of experienced segregation on access to local amenities (instead of 15-minute usage directly). In line with the findings in Figure \ref{fig: psi access income}, we only estimate a positive association between local trips (or local access) and experienced segregation for residents of neighborhoods in the bottom quartile of the income distribution.

\section{Discussion}

Our paper analyses local trip behavior in US cities using GPS data on individual trips from 40 million mobile devices. Using these data, we measure 15-minute usage, which is defined as the share of trips made within a 15-minute walk from home. This measure of local living corresponds to the 15-minute city advocated by urban planners. Urban areas in the US are far from the 15-minute city. The median US city resident makes only 12\% of daily trips to basic amenities locally, although there is considerable regional and socioeconomic variation.

We find a strong correlation between access to stores and amenities within 15-minutes and 15-minute usage.  This relationship is robust to a variety controls. However, the strength of this correlation does not necessarily reflect the causal effect of improving access on local trips. We address this by implementing an instrumental variables approach that leverages variation in access created by the 1961 New York City Zoning ordinance across New York neighborhoods. We use the average maximum commercial floor area ratio (FAR) allowed in each neighborhood by the plan as an instrument for current access to amenities and services. Places with more permissive commercial zoning in 1961 have a larger share of local trips today. We interpret this as evidence supporting the view that local zoning rules shape the level of access to nearby amenities, and that more permissive local zoning provides a natural policy level for those advocates interested in increasing the number of local trips.

In the final part of the paper, we test whether 15-minute access and usage leads to urbanites living more segregated lives. More localized trips and more access to local amenities is associated with higher levels of experienced segregation for low-income residents because the opportunities for higher-income social interactions in most cities are located away from low-income neighborhoods. One consequence of 15-minute cities may be the increased isolation of the poor.  

The findings presented in this paper contribute to our understanding of the mechanisms behind urban travel patterns and underscore the value of mobility data for addressing environmental and social sustainability in cities. Taken together, our results suggest that the planning strategies embodied by the ``15-minute'' city could be effective in reducing trip lengths, but such policies should be balanced with strategies to mitigate potential negative social externalities. Moreover, our work offers a simple empirical framework that could be used for policy evaluation of urban interventions that promote sustainable mobility and shorter trips.

Some limitations and potential future extensions are worth noting. The 15-minute city is built around the idea that everyone should reach their essential social functions within a short sustainable commute. Our usage data is aggregated to the POI level and does not contain information about individuals' work locations. Consequently, our analysis is silent about commuting trips. Further, since our data are aggregated to the neighborhood level, we cannot infer the mode of travel of each person in our data. This means that we can only measure the potential for trips within walking distance, leaving open the question of whether people actually walk to meet their daily needs. Moreover, we only focus on an income-based measure of experienced segregation. Future work could consider segregation by race and ethnicity.  Finally, our causal analysis in New York points towards a causal relationship between amenity access and local usage, but we still cannot control for the selection of people into neighborhoods and this relationship might not be generalized to other US cities. Future work could explore this in more detail for a larger sample of cities.

{\small \renewcommand{\baselinestretch}{1} \addtolength{\parskip}{.65pt}
               \bibliographystyle{econometrica_etal}
               \bibliography{15_minutes.bib}
}
\pagebreak

\setcounter{figure}{0} \renewcommand{\thefigure}{S\arabic{figure}}
\setcounter{table}{0} \renewcommand{\thetable}{S\arabic{table}}
\setcounter{page}{1} \renewcommand{\thepage}{S\arabic{page}}

\section{Methods}

\subsection{Measuring 15-minute usage and access}
\label{app: indices}

To calculate our usage index for a given block group, we first restrict our points of interest to those located within the block group's urban area. We then calculate the total number of trips taken by residents of that block group to all local trips within the block group's 15-minute walkshed and divide it by the total number of local trips taken by residents of that block group. If residents of a given block group take no trips within our data, they are given a usage index value of 0.

In order to aggregate our usage index to the urban area level, we take a population-weighted average as follows:
$$\text{usage}_A = \frac{\sum_{\{a\in A\}} \text{usage}_a * \text{population}_a}{\text{population}_A},$$
where $A$ represents an urban area and $\{a\in A\}$ represent the block groups contained within it.

To calculate our access index for a given block group, we overlay our walksheds with
the point of interest locations in our dataset, identifying the number of amenities in each category that are reachable within a 15-minute walk of each home block group. We then convert each amenity total by category into a percentile across all census block groups, giving a value between 0 and 100 for every block group, for each category. We then construct weights of relative importance for each category by calculating the total percentage of trips to that category in our dataset. The reasoning here is that access to places that people use on a more frequent basis is more important to local living than access to places that are rarely visited. Finally, we combine all of this into an access index for block group $a$ as follows:
$$\text{access}_a = \sum_{\{c \in \text{categories}\}} \text{access percentile}_{a,c} * \text{weight}_c\text{, where}$$
$$\sum_{\{c \in \text{categories}\}}\text{weight}_c = 1$$

Again, in order to aggregate our usage index to the urban area level, we take a population-weighted average as follows:
$$\text{access}_A = \frac{\sum_{\{a\in A\}} \text{access}_a * \text{population}_a}{\text{population}_A},$$
where $A$ represents an urban area and $\{a\in A\}$ represent the block groups contained within it.

\subsection{Supplementary data sources}
\label{app: other data}
For the socioeconomic data, including population, land area (to compute densities), income, and education in our empirical analysis, we use data for 2019 from the American Community Survey 5-year Estimates by census block group for the 2015-2019. The location of transit stops (to construct transit stop density in each census block group) comes from the National Transit Map - Stops dataset which was compiled from the Bureau of Transportation Statistics (BTS), and is part of the U.S. Department of Transportation/BTS National Transportation Atlas Database \citep{BTS}.

To compute the share of trips taken in urban areas for each region, we match each census block group centroid to the region that contains it. We obtain the region geometry from the 2016 census. 

The additional data for New York City used in the IV estimation comes from two sources. To obtain the spatial distribution of restrictions established by the 1961 Zoning Resolution, we combine the information on FAR levels assigned to each type of zoning districts from the \href{https://www1.nyc.gov/assets/planning/download/pdf/about/city-planning-history/zoning_handbook_1961.pdf}{1961 Zoning Handbook} with the map of 1961 zoning districts provided to us by the Department of City Planning. To control for various neighborhood characteristics observed pre-1961 we use the 1960 census data at the census tract level. 

\subsection{Physical Segregation Index}\label{app: seg measure}
For a given individual of income $k_i$ located in point of interest $L$, we calculate the integration, or diversity of interactions, that they experience in that POI as a weighted average socioeconomic difference between them and every other SafeGraph user located in that POI as follows:

$$\text{experienced integration}_{k_i,L} = \frac{\sum_{k_j}p_{k_j, L}\cdot s_{k_i\to k_j}}{\sum_{k_j}p_{k_j, L}},$$
$$s_{k_i\to k_j} = |r_i - r_j|$$
where $p_{k_j,L}$ is the number of people of income $k_j$ who visit $L$ and $r_j$ is the income rank of individuals of income $k_j$. We then aggregate up to the home block group level, describing the experienced integration of individuals from block group $j$ as:
$$\text{experienced integration}_{j} =\frac{\sum_{L \in \text{POIs}}\text{experienced segregation}_{k_j,L}\cdot p_{j,L}}{\sum_{L \in \text{POIs}}p_{j,L}},$$
where $p_{j,L}$ is the number of people from $j$ who visit $L$.

Finally, we define:
$$\text{experienced segregation}_{j} = 1 - \text{experienced integration}_{j} $$
We then scale the measure so that it ranges between 0 and 100, where 100 corresponds to the highest level of experienced segregation and 0 to the lowest. 

\section{Tables and Figures} 
\begin{table}[H] \centering 
  \caption{Correlation of various measures of local living with 15 min. usage} 
  \label{tbl:corr cut-offs} 
\resizebox{\textwidth}{!}{\begin{tabular}{L{3cm}cccc}\toprule
\multicolumn{1}{c}{Alternate measure} & \multicolumn{1}{c}{Neighborhood  level correlation} & \multicolumn{1}{c}{Neighborhood level mean} & \multicolumn{1}{c}{Urban area level correlation} & \multicolumn{1}{c}{Urban area level mean} \\ 
& (1) & (2) & (3) & (4) \\\midrule
\multicolumn{1}{c}{15 min. usage} & 1    & 0.225 & 1 & 0.134 \\ 
\multicolumn{1}{c}{10 min. usage} & 0.901 & 0.172 & 0.957 & 0.106 \\ 
\multicolumn{1}{c}{20 min. usage} & 0.941 & 0.282 & 0.982 & 0.188 \\ 
\multicolumn{1}{c}{25 min. usage} & 0.892 & 0.335 & 0.962 & 0.233 \\ 
\multicolumn{1}{c}{Med. trip dist. (km)} & -0.727 & 2.704 & -0.626 & 3.499 \\ 
\multicolumn{1}{c}{Observations} & 147,760 & 147,760 & 418 & 418 \\ \bottomrule
\end{tabular} }
\begin{minipage}{1\linewidth}
\scriptsize \textsl{Notes.---} The table shows neighborhood level correlations and means for different measures of local usage. Column 1 provides the correlation between our main usage measure and alternative measures defined by the share of trips within 10, 20, and 25 minutes, as well as the median trip length. Column 2 reports (unweighted) means for these alternative variables. Columns 3 and 4 aggregate the data to urban areas and report urban area-level correlations and (unweighted) means.
\end{minipage}

\end{table} 

    \begin{table}[H]
    \centering
    \caption{Comparison of local access metrics}
    \label{access metrics corr}
\resizebox{\textwidth}{!}{\begin{tabular}{lrrrrrrrr}
\toprule
 & &\multicolumn{7}{c}{Counts of Amenities within a 15-minute walk (CBG-level)} \\
\cmidrule(l{3pt}r{3pt}){3-9}
 & Equal-Weighted Access & All & Healthcare & Restaurants & Services & Groceries & Religious & Arts and Culture\\
 & (1) & (2) & (3) &(4) & (5) &(6) & (7) & (8)\\
\midrule
Correlation with Access & 0.98 & 0.51 & 0.37 & 0.45 & 0.52 & 0.60 & 0.59 & 0.30 \\
\bottomrule
\end{tabular}}
\begin{minipage}{1\linewidth}
\scriptsize \textsl{Notes.---} The table shows neighborhood level correlations between different measures of local access to services and amenities. Column one provides the correlation between our main access index and an alternative measure of access, where the scores associated with the number of amenities in each category are added with equal weights. Columns 2--8 provide correlations between our baseline measure and counts of the number of each type of amenity within a 15-minute walkshed.		
\end{minipage}
\end{table}
    \begin{figure}[H]    
        \centering  
        \includegraphics[width=0.7\textwidth]{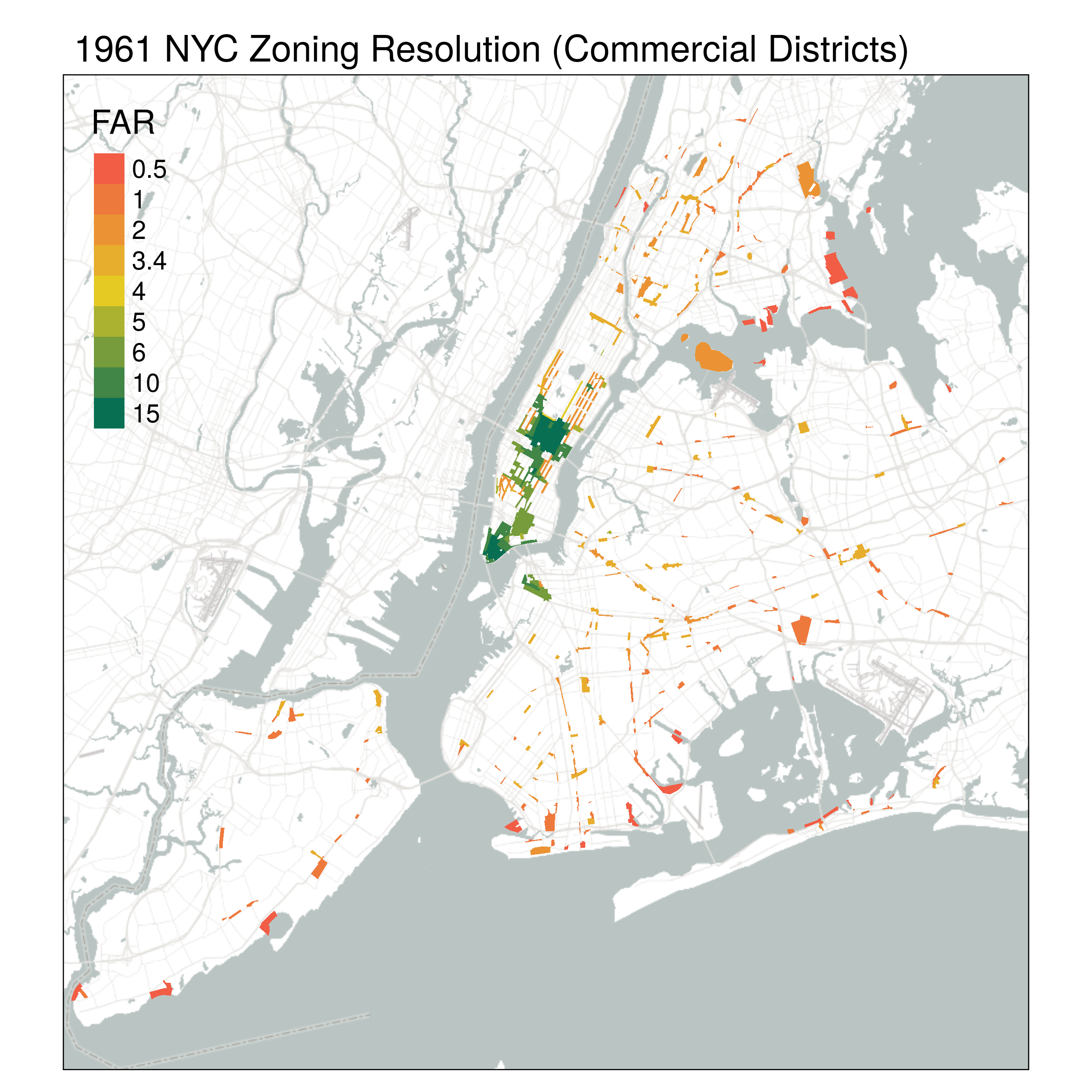}
        \caption{\textsc{Historical zoning.} Maximum floor-to-area ratio in commercial uses established according to the 1961 New York City Zoning Resolution.}
        \label{fig: zoning map}     
    \end{figure} 
\begin{table}[!htbp] \centering 
  \caption{First-stage and reduced-form estimates of 1961 zoning regulation on current access and usage.} 
    \label{regs_far_usage_access}
\resizebox{\textwidth}{!}{
\begin{tabular}{@{\extracolsep{5pt}}lcccc} \toprule
 & \multicolumn{4}{c}{\textsl{Dependent variable:}} \\ 
 & \multicolumn{2}{c}{Access} & \multicolumn{2}{c}{Usage (\% trips)} \\\cmidrule(r){2-3}\cmidrule(r){4-5}
 & (1) & (2) & (3) & (4)\\ 
\midrule 
 Log Average Commercial FAR & 0.927$^{***}$ & 0.548$^{***}$ & 7.185$^{***}$ & 5.872$^{***}$ \\ 
  & (0.026) & (0.029) & (0.607) & (0.710) \\ 
  & & & & \\ 
Controls: &  &  &  &  \\ 
Distance to City Center &  & \checkmark &  & \checkmark \\ 
1960 Population and Housing Density &  & \checkmark &  & \checkmark \\ 
1960 Workers by Industry & & \checkmark & & \checkmark \\ 
Observations & 5,170 & 5,170 & 5,170 & 5,170 \\ 
\bottomrule
\end{tabular} 
} 
	\begin{minipage}{1\linewidth}													\scriptsize \textsl{Notes.---} The table reports first-stage (columns 1 and 2) and reduced-form (columns 3 and 4) estimates of the relationship between 1961 zoning and access and usage for neighborhoods in New York City. Columns 2 and 4 control for distance to the city center and a host of socio-economic controls for 1960, including population and housing density and the share of workers by industry.  The covariates included in these models come from \citet{twinam2017} and \citet{martynov2022}.  Robust standard errors are in parentheses. The coefficients with *** are significant at the 1\% confidence level; with ** are significant at the 5\% confidence level; and with * are significant at the 10\% confidence level.
	\end{minipage}	
\end{table}     
\begin{table}[!htbp] \centering 
  \caption{Relationship between experienced segregation and local trips for neighborhoods of different income levels.} 
  \label{tbl : psi usage income} 
\resizebox{\textwidth}{!}{\begin{tabular}{@{\extracolsep{5pt}}lcccc} 
\toprule
&\multicolumn{4}{c}{\textit{Dependent variable:}} \\ &\multicolumn{4}{c}{Physical Segregation Index (PSI)} \\ 
& \multicolumn{2}{c}{Main 15 urban areas} & \multicolumn{2}{c}{Full sample} \\ \cmidrule(r){2-3}\cmidrule(r){4-5}
 & (1) & (2) & (3) & (4)\\\midrule
Usage $\times$ Low income &       0.165$^{***}$&                    &       0.070$^{***}$&                    \\
            &     (0.029)        &                    &     (0.013)        &                    \\
Usage $\times$ Moderate-low income &       0.024        &                    &      -0.002        &                    \\
            &     (0.015)        &                    &     (0.007)        &                    \\
Usage $\times$ Moderate-high income &      -0.081$^{***}$&                    &      -0.064$^{***}$&                    \\
            &     (0.017)        &                    &     (0.007)        &                    \\
Usage $\times$ High income &      -0.113$^{***}$&                    &      -0.064$^{***}$&                    \\
            &     (0.022)        &                    &     (0.010)        &                    \\
Access $\times$ Low income &                    &       0.182$^{***}$&                    &       0.039$^{***}$\\
            &                    &     (0.041)        &                    &     (0.014)        \\
Access $\times$ Moderate-low income &                    &       0.024$^{*}$  &                    &      -0.019$^{***}$\\
            &                    &     (0.014)        &                    &     (0.006)        \\
Access $\times$ Moderate-high income &                    &      -0.096$^{***}$&                    &      -0.076$^{***}$\\
            &                    &     (0.014)        &                    &     (0.007)        \\
Access $\times$ High income &                    &      -0.119$^{***}$&                    &      -0.071$^{***}$\\
            &                    &     (0.015)        &                    &     (0.010)        \\
\\ Urban area fixed effects & \checkmark & \checkmark & \checkmark & \checkmark \\ R-squared&        0.26        &        0.28        &        0.34        &        0.35        \\
Observations&       34594        &       34594        &      143411        &      143411        \\
\bottomrule
\end{tabular}} 

\begin{minipage}{1\linewidth}
	\scriptsize \textsl{Notes.---} The table reports OLS estimates of the relationship between experienced segregation and 15-minute usage (columns 1 and 3) and access (columns 2 and 4). We provide estimates for neighborhoods in the select sample of cities from Panel C of Figure \ref{fig: usage map} and for all neighborhoods in our sample. The table reports estimates interacted with dummies for low income neighborhoods (bottom quartile of income in urban area), moderate-low income neighborhoods (second quartile),  moderate-high income neighborhoods (third quartile), and high income neighborhoods (top quartile).   
Robust standard errors clustering at the urban area level are in parentheses. The coefficients with *** are significant at the 1\% confidence level; with ** are significant at the 5\% confidence level; and with * are significant at the 10\% confidence level.
\end{minipage}
\label{a6}
\end{table}   
\end{document}